\documentclass[a4paper,11pt]{article}
\usepackage{pos}

\title{The SABRE South Experiment at the Stawell Underground Physics Laboratory}
\ShortTitle{The SABRE South Experiment}

\author*[a,b]{L.~J.~Milligan on behalf of the SABRE South Collaboration}

\affiliation[a]{School of Physics, The University of Melbourne,\\ Parkville, VIC 3010, Australia}

\affiliation[b]{ARC Centre of Excellence for Dark Matter Particle Physics, Australia}

\emailAdd{lmilligan@student.unimelb.edu.au}

\abstract{SABRE aims to provide a test of the signal observed by DAMA/LIBRA through two separate detectors that rely on joint ultra-high NaI(Tl) purity crystal R\&D activities: SABRE South at SUPL Australia and SABRE North at LNGS Italy. SABRE South is designed to disentangle seasonal and site-related effects from the dark matter-like modulated signal. Ultra-high purity crystals are immersed in a liquid scintillator veto, further surrounded by passive shielding and a plastic scintillator muon veto. Significant work has been undertaken to assess and mitigate background from the detector materials, and to understand the performance of both the crystal and veto systems. SUPL is a newly built facility located 1024 m underground in Australia. SABRE South is currently being assembled and will be completed in 2025, with first subsystems already installed in SUPL. This proceedings will report on the general status of the SABRE South assembly, its expected performance, and the design of SUPL.}

\FullConference{42nd International Conference on High Energy Physics (ICHEP2024)\\
18-24 July 2024\\
Prague, Czech Republic\\}

\begin{document}
\maketitle

\section{Introduction}

 To date, the only reported direct detection of dark matter (DM) has come from the DAMA/LIBRA experiment, which claims to observe an annually modulating nuclear recoil rate with a NaI(Tl) target. This annual modulation is expected to arise from the relative motion of the Earth around the Sun, and the Sun around the galactic centre, such that a scattering rate increases as the Earth moves into the DM halo, and vice-versa as it moves away. DAMA/LIBRA has observed their reported signal for 20 years in the 2-6~keV$_{\rm ee}$ region at a significance of 12.4$\sigma$, and recently also in the 1-6~keV$_{\rm ee}$ region at 11.6$\sigma$~\cite{Bernabei:2023gsv}. Modulating annually, the reported signal has a period of 1~year ($t_0 = 152.5$~days) that peaks on June 2nd with a very low modulation amplitude of $\sim$0.01~cpd/kg/keV$_{\rm ee}$. This result remains unverified, and is in tension with limits imposed by other direct detection experiments (such as LXe based detectors) in the same mass regime~\cite{PPDG2018}. To reliably test this result in a model-independent way, any such detector would need to use the same target material.
 
 The SABRE South experiment is the Southern Hemisphere component of the SABRE Experiment, which aims to test the modulating signal observed by the DAMA/LIBRA experiment using ultra-pure NaI(Tl). Being located in the Southern Hemisphere uniquely allows the disentanglement of seasonal and site-related effects from the DM-like modulated signal, due to the seasonal shift relative to the Northern Hemisphere. The SABRE South detector will be located in the Stawell Underground Physics Laboratory (SUPL), which is the first underground laboratory in the Southern Hemisphere. Installation of SABRE South subsystems in SUPL has already begun, and assembly of the full detector is expected to be completed in 2025.

\section{The SABRE South Detector}
\label{sec:det}

\begin{figure}[htb]
    \centering
    \includegraphics[width=0.59\linewidth]{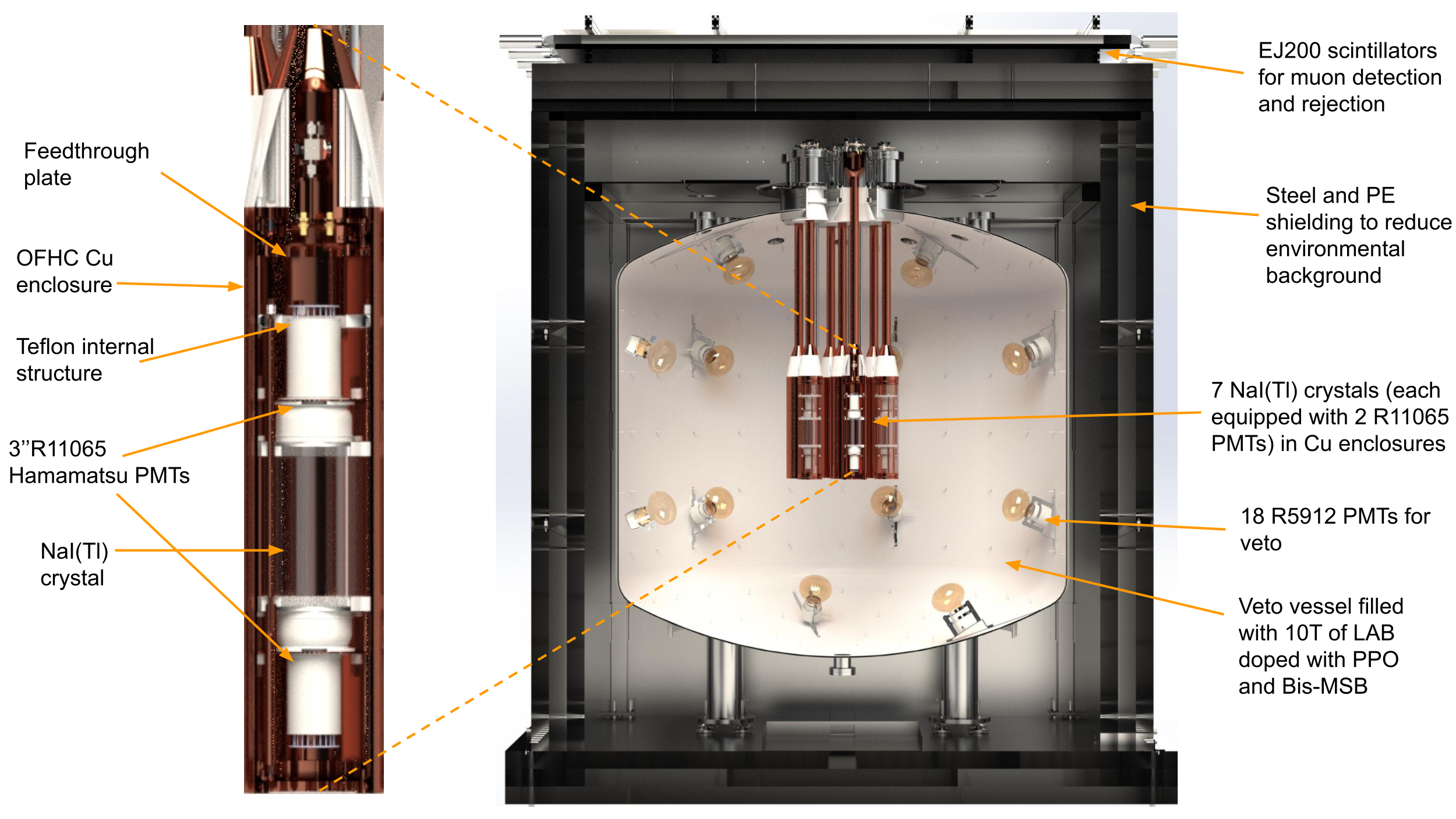}
    \includegraphics[width=0.4\linewidth]{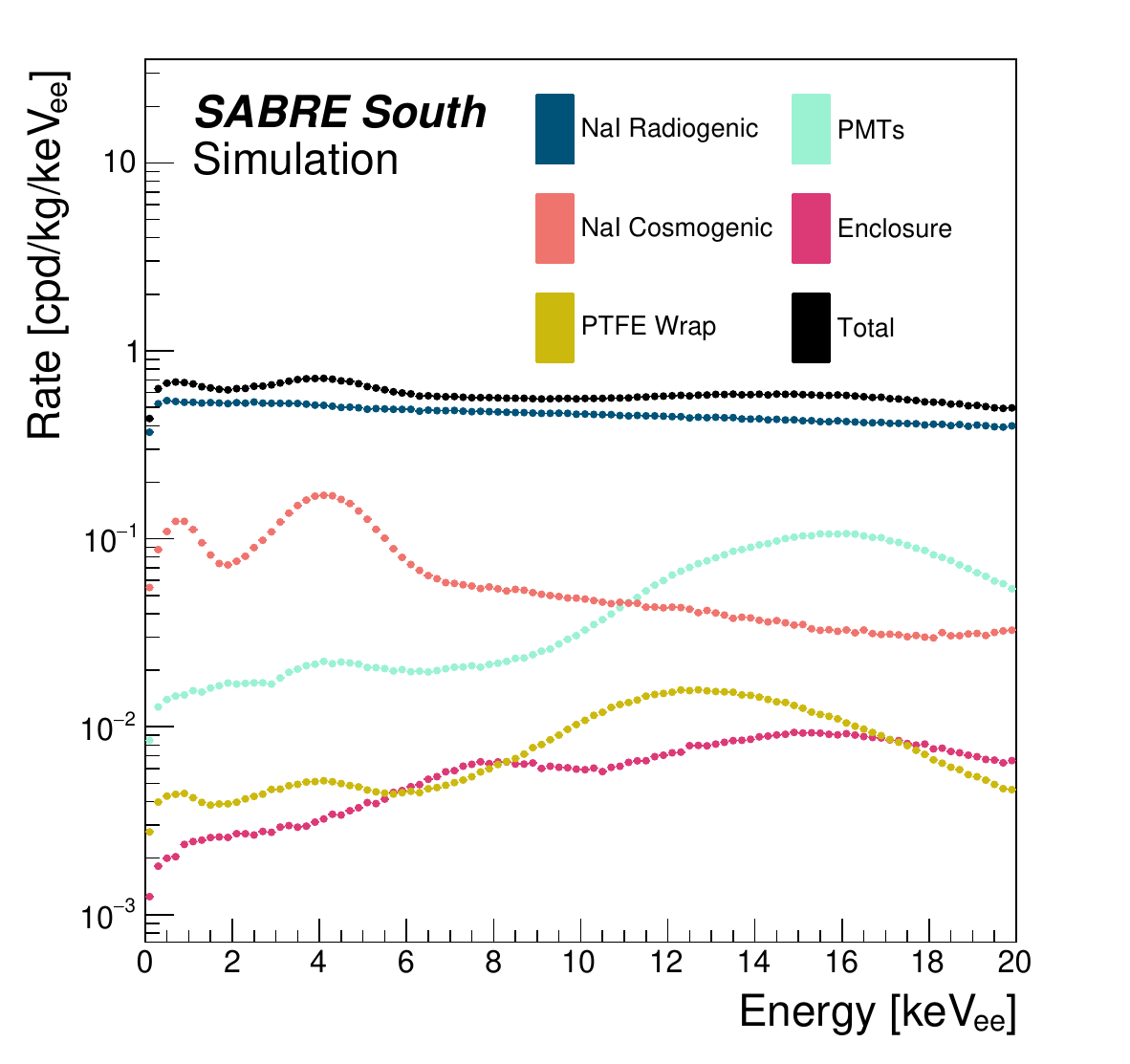}
    \caption{\textbf{Left:} A labelled render of the SABRE South detector system, including shielding. Left of the detector is an enlarged version of the crystal enclosure, labelling its key components. \textbf{Right:} The total experimental background for SABRE South as a function of energy.}
    \label{fig:detector}
\end{figure}

The SABRE South detector is composed of three subdetectors: an array of seven ultra-pure NaI(Tl) crystals enclosed in oxygen free high thermal conductivity (OFHC) copper enclosures, a linear alkylbenzene (LAB) based liquid scintillator (LS) veto, and a 9.6~m$^2$ dedicated muon veto. Surrounding the detector is passive steel and high density polyethylene shielding, in which the latter is sandwiched between the former. Figure~\ref{fig:detector} depicts the full detector system. 

The seven crystals will sum to a total mass of between 35 to 50~kg, depending on whether the crystals are grown with SICCAS by the seedless Bridgman method ($7\times7$~kg crystals) or by RMD using a combination of the vertical Bridgman method and a zone-refining technique ($7\times5$~kg crystals). As we expect the smaller crystals to have a lower background rate due to zone-refining, the expected sensitivity of SABRE South should not be drastically changed~\cite{TDRFull}. Test crystals have been measured to have extremely high purity levels~\cite{Antonello:2020xhj}. A single crystal enclosure contains one ultra-pure crystal directly coupled to two Hamamatsu 7.62~cm R11065 photomultiplier tubes (PMTs), and is continuously purged with high purity nitrogen. These PMTs have been rigorously pre-calibrated, so to understand the future performance of each detector. The enclosure is suspended in the LS by an OFHC copper conduit. A remotely operated radioactive calibration system for the crystal arrays has been designed~\cite{TDRFull}. The crystal detectors are expected to have a 1~keV$_{\rm ee}$ threshold, such that the 1-6~keV$_{\rm ee}$ region of interest can be probed.

The active LS veto is a key feature of the SABRE South detector. Its main requirements are to veto extrinsic and key intrinsic crystal backgrounds, particularly the $^{40}$K impurity which decays an Auger electron right in the 1-6~kev$_{\rm ee}$ region of interest, and a gamma ray that can be tagged by the veto. Simulations of the SABRE South experimental background indicate that the $^{40}$K background in the crystal can be reduced by a factor of $\sim$10 by the LS veto, with a veto efficiency of 27\% with respect to the total experimental background --- meaning $<10$\% of the total background is external to the crystals~\cite{sabre_background}. This subdetector can also be used for background characterisation, both particle identification and position reconstruction, providing an avenue for understanding what type of background processes may produce a DAMA/LIBRA like modulation. The LS veto is composed of 12,000~L of LAB sourced from the supplier to the JUNO experiment~\cite{JUNO:2021vlw}, doped with approximately 3.5~g/L of PPO and 15~mg/L of bis-MSB. At least 18 oilproof Hamamatsu 20.2~cm R5912 PMTs will be installed along the walls of the vessel, which will be covered in reflective PET Lumirror sheets to increase photon detection efficiency. This could be increased to 32, augmented with decommissioned PMTs from the Daya Bay experiment, thus improving the photosensor coverage of the detector and the effectiveness of reconstruction techniques. Preliminary quality assurance tests have been performed on the 16 PMTs receieved from Daya Bay, with current results indicating 14 are suitable for use in the LS veto, whilst the remaining 18 have been undergone pre-calibration tests. A threshold of 50~keV is possible for this subdetector, which is equal to approximately 10 photoelectrons (PEs), where each PMT is expected to be detecting 1-2~PEs. As a result, thorough pre-calibration of all R5912 PMTs to be used in this system is imperative. The LS veto is calibrated remotely with a radioactive calibration system, and an optical calibration system that allows for remote characterisation/monitoring of PMT properties throughout the lifetime of the experiment. 

The muon veto sits above the shielding of the detector, and consists of eight panels of EJ-200 plastic scintillator with two Hamamatsu R13089 PMTs coupled to either end. This detector can be used in coincidence with the LS veto to improve background reconstruction, but is also capable of position reconstruction along the length of each panel to a resolution of 5~cm. To calibrate this detector a linear stage is placed over the width of the detector array, with a $^{60}$Co source nominally positioned at the centre of each of the 8 panels for remote calibration. This detector system is currently installed in SUPL and arranged in telescope mode for flux measurements.

The total experimental background of the SABRE South detector is expected to be 0.72~cpd/kg/keV$_{\rm ee}$~\cite{sabre_background}. Figure~\ref{fig:detector} shows the total experimental background as a function of energy, with the veto applied. 

\section{The Stawell Underground Physics Laboratory}

\begin{figure}[htb]
    \centering
    \includegraphics[width=0.4\linewidth]{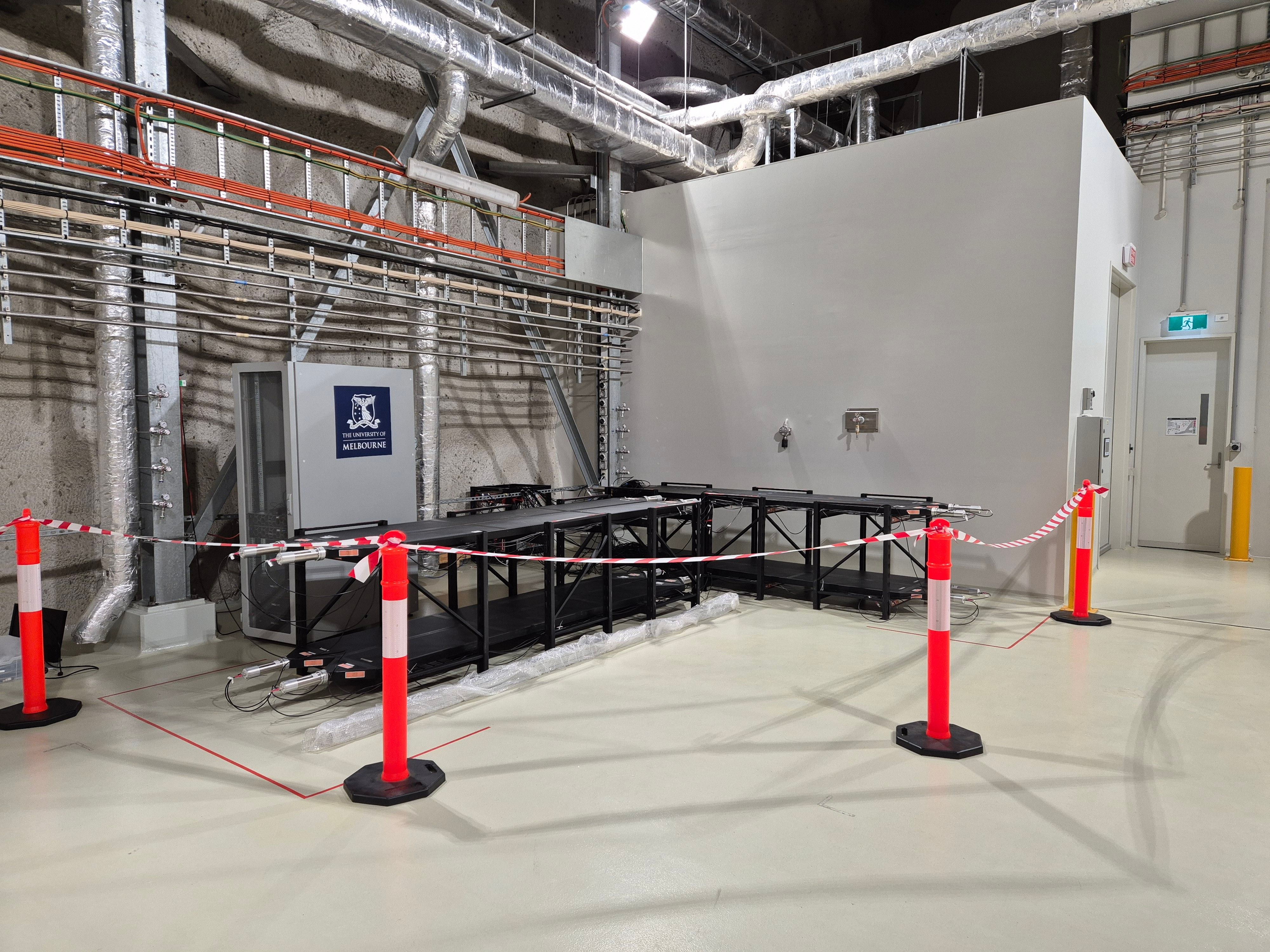}
    \includegraphics[width=0.5\linewidth]{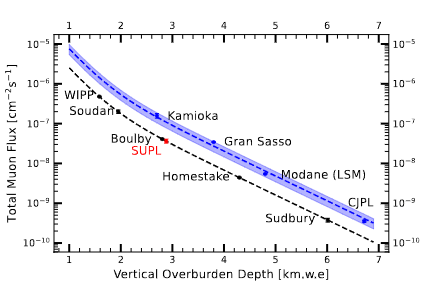}
    \caption{\textbf{Left:} The SABRE South detectors assembled for flux measurements in SUPL. \textbf{Right:} The total muon flux as a function of water equivalent depth. The muon flux for SUPL is relatively on par with Boulby and Kamioka. Labs with mountainous overburden are shown in the blue shaded region.}
    \label{fig:muonsSUPL}
\end{figure}

The Stawell Underground Physics Laboratory (SUPL) is the first underground physics lab in the Southern Hemisphere. SUPL is 1024~m deep, with a flat overburden, and has been constructed in the Stawell Gold Mine, in Stawell, Victoria, Australia (approximately 240~km north west of Melbourne). The lab is shielded by $\sim$2900~km of water equivalent shielding, which suppresses the muon flux to $\sim3\times10^{-8}$~cm$^{-2}$s$^{-1}$, shown in Fig.~\ref{fig:muonsSUPL}. The lab was completed for first access in 2024. The commissioning of the first detector occurred in early 2024, as the SABRE South muon veto was assembled in ``telescope mode" for measurements of the muon flux and the angular spectra, as seen in Fig.~\ref{fig:muonsSUPL}. 

Data is currently being collected with this setup, with analysis ongoing, and it has provided the first test of the remote data acquisition system and processing pipeline. The results from the analysis will not only serve as validation of our current simulated flux/angular spectrum, but can also be used in combination with data collected after full detector commissioning to measure the annual muon flux modulation. These measurements will be the first of their kind in SUPL, and are expected to be public by the end of 2024.


\section{Photomultiplier Pre-calibration Campaign}

The PMTs for both the crystal detectors and the LS veto have been undergoing rigorous pre-calibration. This is crucial, as both subdetectors have requirements and features that necessitate a deep understanding of their PMTs. In the case of the crystal detectors, understanding of the noise, gain, and single photoelectron (SPE) response the R11065 PMTs is essential to ensure the 1~keV$_{\rm ee}$ threshold is reached. Furthermore, since other experiments demonstrate unexplained backgrounds between 1-2~keV$_{\rm ee}$~\cite{Coarasa:2024xec}, an understanding of the PMT noise rate is essential, as well as the development of techniques that can suppress it. For the LS veto, given the low yield of detectable photons for a single PMT, such that the probability of detecting a single PE is $\sim$0.20~PE/keV, understanding the SPE response/gain and noise of each PMT is essential for appropriate SPE thresholds and gains to be set for each PMT. This is to ensure that we can hit the 50~keV threshold desired for optimal veto efficiency and reconstruction capabilities, whilst also controlling for the noise rate in the detector. 

All R5912 veto PMTs have been calibrated (i.e. all 20+16 SABRE South R5912 PMTs) for some key properties: SPE response and gain, timing characteristics, and dark rate as a function of temperature. Other properties such as response linearity/saturation, relative quantum efficiency, afterpulsing, and spontaneous light emission from the epoxy oilproof base, have also been measured. The measurements of these additional properties is motivated by reconstruction performance in the detector, as well as understanding all noise contributions.

For the R11065 PMTs, measurements of the SPE response, gain, dark rate as a function of temperature, afterpulsing, dynode glow, and timing properties are ongoing. We plan to perform these measurements on the full set of R11065 PMTs, so that each crystal detector can be well understood. In addition to calibration measurements, a noise vs. scintillation signal classifier (utilising a boosted decision tree) has been developed with using a commercial EPICS NaI(Tl) crystal, with the aim of demonstrating a method for reducing noise-related backgrounds between 1-2~keV$_{\rm ee}$, and potentially reducing our experimental threshold below 1~keV$_{\rm ee}$. This is motivated by background between 1-2~keV$_{\rm ee}$ seen by other experiments~\cite{Coarasa:2024xec}. 

Publications detailing the pre-calibration of both PMT types will be public by the end of 2024.

\section{Physics Program}

The SABRE South collaboration is exploring different types of signals and DM models that our detector can measure --- leveraging both the crystal detectors, and the LS veto detector. Preliminary sensitivity studies have been performed for the Migdal effect, bosonic super-WIMPs, and supernova neutrinos~\cite{TDRFull}. Results from the supernova neutrino sensitivity study indicate that the LS veto is sensitive to neutrinos from supernovae occurring at any distance up to the galactic centre. As a result, there is the possibility for SABRE South to join a supernova early warning system. 

Additionally, following the release of a study suggesting that the annual modulation reported by DAMA/LIBRA could be induced by their background subtraction technique~\cite{COSINE-100:2022dvc}, SABRE South has conducted our own study using a best faith reproduction of DAMA/LIBRA's background. Our study demonstrates that the DAMA/LIBRA background is low enough, given their over-estimated tritium activity, that the background shape and subtraction method cannot induce a modulation \cite{James:2024gtl}. Furthermore, any induced modulation is opposite in phase to the signal DAMA/LIBRA reports. This result is in contrast to what was previously suggested in Ref.~\cite{COSINE-100:2022dvc}.

\section{Conclusion}

\begin{figure}[htb]
    \centering
    \includegraphics[width=0.4\linewidth]{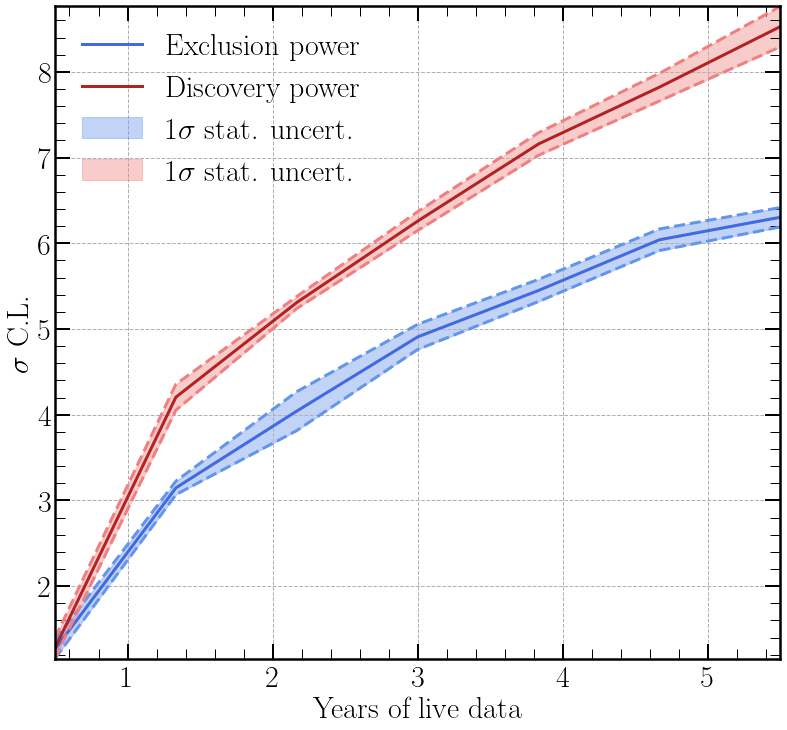}
    \includegraphics[width=0.45\linewidth]{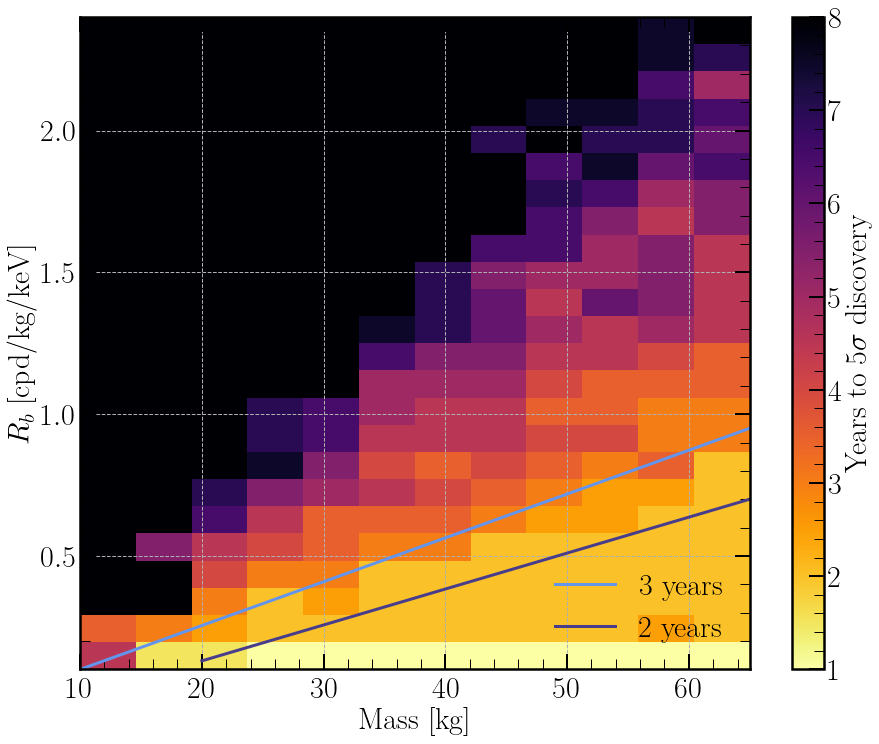}
    \caption{\textbf{Left:} Exclusion/discovery power of SABRE South given 50~kg of crystals and a total background of 0.72~cpd/kg/keV$_{\rm ee}$. \textbf{Right:} Years to 5$\sigma$ discovery as a function of target mass and total background, where the same discovery power is possible with a lower target mass given a lower total background, as discussed in Sec.~\ref{sec:det}.}
    \label{fig:sensitivity}
\end{figure}

The SABRE South detector will be assembled and commissioned in 2025. SUPL is built and the SABRE South muon detectors are already commissioned underground. As shown in Fig.~\ref{fig:sensitivity}, with a total experimental background of 0.72~cpd/kg/keV$_{\rm ee}$ and a 50~kg target mass, SABRE South is capable of 5$\sigma$ discovery (3$\sigma$ exclusion) power to a DAMA-like signal with 2 years of data. 

\bibliographystyle{JHEP}
\bibliography{refs.bib}

\end{document}